\documentclass{aastex}          
\usepackage{spr-astr-addons}    
\usepackage{amsmath}
\usepackage{amssymb}

\def\note #1]{{\bf #1]}}
\def\vec#1{\ensuremath{\mathchoice{\mbox{\boldmath$\displaystyle#1$}}
{\mbox{\boldmath$\textstyle#1$}}
{\mbox{\boldmath$\scriptstyle#1$}}
{\mbox{\boldmath$\scriptscriptstyle#1$}}}}

\begin{document}
%
\title{Stellar turbulence and mode physics}

\shorttitle{Stellar turbulence and mode physics}
\shortauthors{< G{\"u}nter Houdek>}

\author{G\"{u}nter Houdek} 
\affil{Institute of Astronomy, University of Vienna, 1180 Vienna, Austria}
\email{guenter.houdek@univie.ac.at} 


\begin{abstract}
An overview of selected topical problems on modelling 
oscillation properties in solar-like stars is presented. 
High-quality oscillation data
from both space-borne intensity observations and ground-based spectroscopic 
measurements provide first tests of the still-ill-understood, superficial
layers in distant stars. Emphasis will be given to modelling the pulsation
dynamics of the stellar surface layers, the stochastic excitation
processes and the associated dynamics of the turbulent fluxes of heat 
and momentum. 
\end{abstract}


%
\section{Introduction}\label{s:intro}
With the high-quality photometric data from the Kepler satellite 
\citep[e.g.][]{Christ07} and spectroscopic data from ground-based 
observation campaigns, and later also from the Danish SONG network 
\citep{Grundahl07}, we shall be able to address more carefully many 
of the current problems in modelling pulsation properties in 
solar-like stars. In solar-like stars the p-mode lifetimes and
amplitudes are crucially affected by the processes that take
place in the outer convectively unstable stellar layers. A proper
modelling of the dynamics of the convective heat and momentum transfer 
is therefore essential. Recent 3D numerical simulations of the largest 
scales of the convection \citep[e.g.][]{SteinNordlund01, Samadi03, 
Georgobiani04, Stein04, Jacoutot08, Miesch08} have been proven to 
be extremely useful for calibrating and testing semi-analytical models 
for convection and stochastic excitation. However, the high-Reynolds-number 
(and low-Prandtl-number) turbulent convection in stars still
prohibits today's simulations from resolving all the required scales of 
stellar turbulence. Consequently such simulations have to use
sub-grid-scale models, which may lead to different results
\citep[e.g.][]{Jacoutot08}. Because this situation will not change in 
the near future, we still need analytical models for describing the 
convection and pulsation dynamics in stars.

Sections 2 and 3 will discuss selected problems of our current understanding of
nonadiabatic pulsation dynamics in the Sun and in the hotter F5 star Procyon. 
Although we can reasonably well reproduce the observed pulsation properties in 
the Sun, the recent Procyon observations 
\citep[e.g.][and references therein]{Arentoft08} have revealed serious 
problems in our models for estimating the oscillation amplitudes in stars 
hotter than the Sun. In Section~4, therefore, I shall address the problem
of selecting a proper temporal turbulence spectrum for modelling the
stochastic energy supply rate for acoustic modes.

\section{Mode parameters}\label{s:modeparam}
In solar-like stars all possible modes of oscillation are 
stable; thus, if a given oscillation mode is somehow excited, its amplitude
will decay
over a finite time, typically of the order of days to months, the inverse of 
which is the damping rate $\eta$. The oscillation power spectrum can 
be described in terms of an ensemble of intrinsically damped, stochastically 
driven, simple-harmonic oscillators, provided that the background equilibrium 
state of the star were independent of time. In that case the mode profile is 
essentially Lorentzian, and the intrinsic damping rates of the modes
could then be determined observationally from 
measurements of the pulsation linewidths.
The other fundamental property of the observed oscillation power spectrum is 
the height, $H$, of a single peak in the Fourier spectrum.
The observed velocity signal $v(t)={\rm d}\xi/{\rm d}t$ (where $\xi(t)$ is 
the surface displacement of the damped, stochastically driven, harmonic 
oscillator, and $t$ is time) can then be related to the mode height $H$ 
by taking the Fourier transformation of the harmonic oscillator
followed by an integration over frequency to obtain the total mean energy $E$
in a particular pulsation mode with inertia $I$ \citep[e.g.][]{Chaplin05, 
Houdek06}. For $T\eta\gg1$,where $T$ is observing time, the squared surface 
rms velocity is then given by

\begin{figure}[t]
\centering
\includegraphics[width=0.45\textwidth]{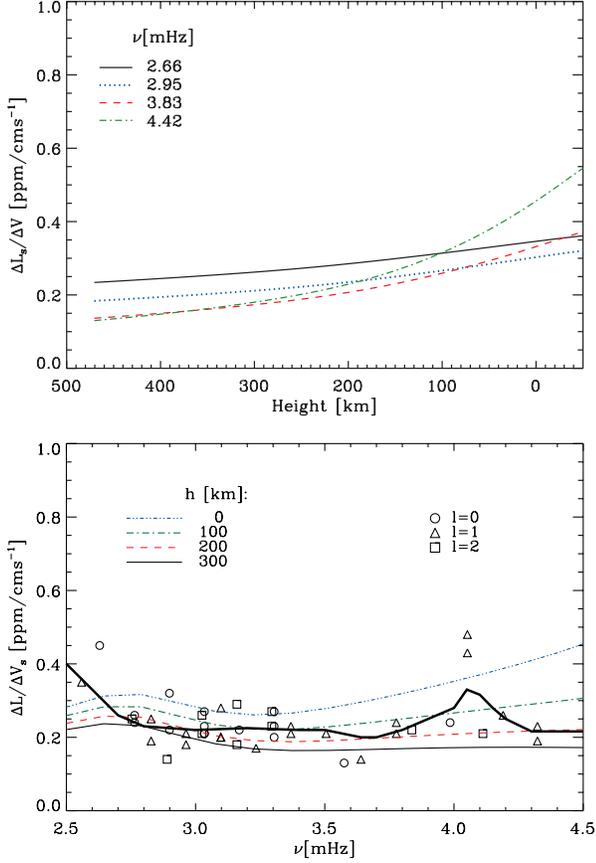}
\caption{
{\bf Top}:
Calculated amplitude ratios (see equation (\ref{e:amprat})) as a function
of height in a solar model for modes with different frequency values.
{\bf Bottom}:
Theoretical amplitude ratios (surface luminosity perturbation over velocity)
for a solar model compared with observations by \citet{Schrijver91}.
Computed results are depicted at different heights above the photosphere
($h$=0\,km at $T=T_{\rm eff}$). The thick, solid curve indicates a
running-mean average of the data \citep[from][]{Houdek99}
}
\label{fig:amprat_sun}
\end{figure}
\begin{figure}[t]
\centering
\includegraphics[width=0.45\textwidth]{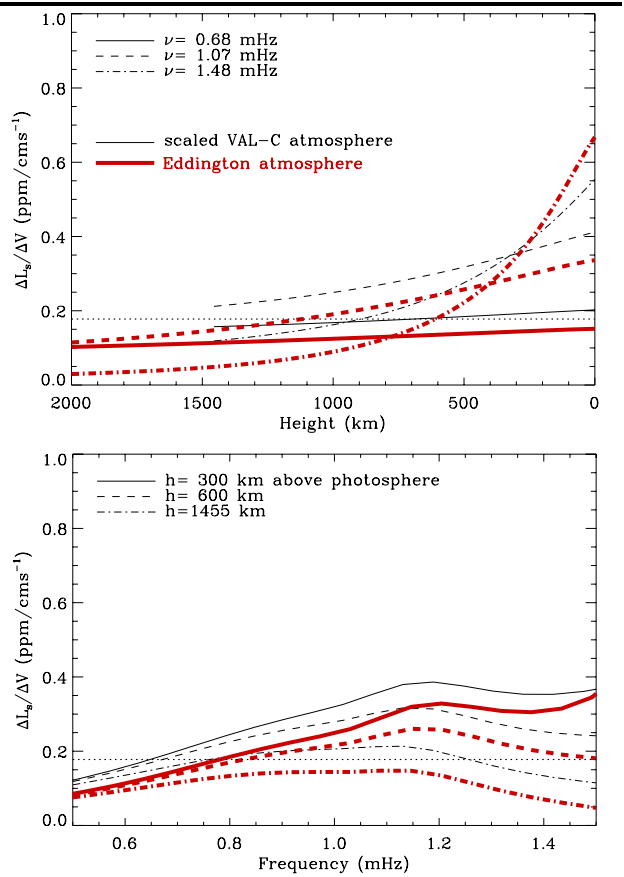}
\caption{Calculated amplitude ratios (see Eq.\,(\ref{e:amprat})) 
for a model of Procyon A are compared with observations
by Arentoft\,(2008; horizontal dotted line). Theoretical 
results are shown for a scaled VAL-C atmosphere (black, thin curves) and 
for an Eddington atmosphere (red, thick curves).
{\bf Top}: 
The theoretical amplitude ratios are shown as a function of height and
for three different pulsation modes. The frequencies of the three 
pulsation modes are indicated.
{\bf Bottom}:
The theoretical amplitude ratios are shown as a function of frequency
at three different heights in the stellar atmosphere. The heights 
above the photosphere $h=0\,$km are indicated}
\label{fig:amprat_procyon}
\end{figure}

\begin{equation}
V^2\,:=\frac{E}{I}\,=\,\frac{P}{2\eta\,I}
\,=\,\frac{1}{2}\eta\,H\,,
\label{e:V-H}
\end{equation}
where $P$ is the energy supply rate in erg$\,$s$^{-1}$, and $H$ is given 
in units of cm$^2\,$s$^{-2}$Hz$^{-1}$. The height $H$ is the maximum of 
the discrete power, and is obtained from integrating the power 
spectral density over a frequency bin:
\begin{equation}
H=\int_{\nu-\hat\delta/2}^{\nu+\hat\delta/2}\vert\hat V(\nu)\vert^2\,{\rm d}\nu\,,
\label{e:H}
\end{equation}
where $\hat V(\nu)$ is the Fourier transform of $v(t)$, $\nu$ is cyclic 
frequency, and $\hat \delta=1/2T$, is the frequency bin determined by 
the observation time $T$.
It is therefore not the total integrated power, $V^2$, that is 
observed directly, but rather the power spectral density 
\citep{Chaplin05}.

\section{Pulsation amplitude ratios}\label{s:amprat}

Linearized pulsation equations for nonadiabatic radial oscillations
can be presented as \citep[e.g.][]{Balmforth92a}:
\begin{eqnarray}
\frac{\partial}{\partial m}\left(\frac{\delta p}{p}\right)&=&
 f\left(\frac{\delta r}{r},\frac{\delta T}{T},\frac{\delta p}{p},\frac{\delta p_{\rm t}}{p},\frac{\delta\Phi}{\Phi}\right)\,,\cr
\frac{\partial}{\partial m}\left(\frac{\delta r}{r}\right)&=&
-\frac{1}{4\pi r^3\rho}\left(3\frac{\delta r}{r}+\frac{\delta\rho}{\rho}\right)\,,\cr
\frac{\partial}{\partial m}\left(\frac{\delta L}{L}\right)&=&
-{\rm i}\omega_i\frac{c_pT}{L}\left(\frac{\delta T}{T}-\nabla_{\rm ad}\frac{\delta p}{p}\right)\,,
\label{eq:pulsation}
\end{eqnarray}
where $\delta$ is the Lagrangian perturbation operator, and for simplicity the
right hand side of the perturbed momentum equation is formally expressed by the
function~$f$. Equations~(\ref{eq:pulsation}), together with an equation 
relating the heat flux to the stratification of the star (the full set of 
equations can be found in, e.g., Balmforth 1992a), are solved subject to 
boundary conditions to obtain the eigenfunctions and the complex angular 
eigenfrequency
$\omega_i=\omega_{{\rm r}i}+{\rm i}\eta_i$ of the mode $i$, where 
$\omega_{{\rm r}i}$ is the (real)
pulsation frequency and $\eta_i$ the linear damping rate in (s$^{-1}$).
The turbulent flux perturbations of heat and momentum, $\delta L_{\rm c}$ and
$\delta p_{\rm t}$, and the fluctuating anisotropy factor $\delta\Phi$ are
obtained from the nonlocal, time-dependent convection formulation
by \citet{Gough77a, Gough77b}.

From the linearized nonadiabatic pulsation equations 
(\ref{eq:pulsation}) theoretical intensity-velocity amplitude ratios
\begin{equation}
\frac{\Delta L_{\rm s}}{\Delta V}:=
                        \frac{\delta L/L}{\omega_{{\rm r}i} r\;\delta r/r}
\label{e:amprat}
\end{equation}
can be compared with observations, without the need of a specific excitation 
model and all its uncertainties in describing the turbulence spectrum.

In the top panel of Fig.~\ref{fig:amprat_sun} the theoretical amplitude ratios
(equation~(\ref{e:amprat})) of a solar model are plotted as a function of 
height for several radial pulsation modes. The square root of the mode kinetic
energy per unit increment of $r$, which is proportional to
$r\rho^{1/2}\delta r$, increases rather slowly with
height; the density $\rho$, however, decreases very rapidly and consequently
the displacement eigenfunction $\delta r$ increases with height. This leads
to the results shown in the upper panel of Fig.~\ref{fig:amprat_sun}. The
decrease in the amplitude ratios with height is particularly pronounced for
high-order modes for which the eigenfunctions vary rapidly in the evanescent
outer layers of the atmosphere. It is for that reason why solar velocity
amplitudes from e.g., the GOLF instrument have larger values than the
measurements from the BiSON instrument (by about 25\%, \citealt{Kjeldsen05}).\\
The lower panel of Fig.~\ref{fig:amprat_sun} compares the estimated solar 
amplitude
ratios (curves) with observed ratios (symbols) as a function of frequency. The
model results are depicted for velocity amplitudes computed at different
atmospheric levels. The observations are obtained from accurate irradiance
measurements from the IPHIR instrument of the PHOBOS 2 spacecraft with
contemporaneous low-degree velocity data from the BiSON instrument at
Tenerife \citep{Schrijver91}. The thick solid curve represents a
running-mean average, with a width of 300$\,\mu$Hz, of the observational data.
The theoretical ratios for $h=200\,$km (dashed curve) show reasonable
agreement with the observations.

In Fig.~\ref{fig:amprat_procyon} model results for the F5 star Procyon A are 
compared with observations (horizontal dotted line) by 
\citet{Arentoft08}. 
Theoretical results are shown for two stellar atmospheres: a VAL-C 
\citep{Vernazza81} atmosphere scaled with the model's effective
temperature $T_{\rm eff}$ (thin curves), and an Eddington atmosphere
(thick curves). For both stellar atmospheres the agreement with the 
observations is less satisfactory than in the solar case, indicating 
that we may not represent correctly the shape of the pulsation eigenfunctions.
Consequently there is need for adopting more realistically computed 
atmospheres in the equilibrium models, particularly for stars with much higher
surface temperatures than the Sun. It should, however, be mentioned that the
current photometric observations are still uncertain.

\section{Stochastic excitation and turbulent spectra}\label{s:excitation}
In solar-like stars the oscillations are driven by the vigorous
turbulent convection in the very outer surface layers. The turbulent
fluid motion generates acoustic waves in a broad frequency range, which
excite a large number of global p modes (of the order of 
10 million p modes in the solar case), and possibly also global
g modes \citep[e.g.][]{Appourchaux09}. In the past several 
stochastic excitation models were proposed 
\citep[for recent reviews see, e.g.][]{Houdek06, Appourchaux09}.
In general all reported excitation models reproduce reasonably well 
the observed oscillation amplitudes for solar-like stars that are
similar or cooler than the Sun. However, for stars that are somewhat 
hotter than the Sun, theoretical predictions overestimate the pulsation 
amplitudes by up to a factor of about four, such as for the F5 star Procyon A.
Additional to the uncertainties in modelling the shape of the pulsation
eigenfunctions in the atmospheric stellar layers (see Section~\ref{s:amprat}), 
modelling of the mode damping rates \citep[see e.g., ][]{Houdek06}
and the turbulent energy spectrum play also a crucial role in estimating 
the mode amplitudes. In this paper I shall only address the
problem of modelling the turbulent spectrum for estimating
the energy supply rate $P$.

\subsection{Stochastic excitation model}\label{ss:excitation}
A review on this topic was recently given by \citet{Appourchaux09}.
Therefore I shall summarize only the most important matters.
According to Eq.\,(\ref{e:V-H}) we need to model the (linear) damping 
rate $\eta$ and the energy supply rate $P$ for estimating the mode height $H$.
In this paper I shall address only the modelling of the energy supply rate.
One of the problems that one faces in deriving a stochastic excitation model
is the separation of the total fluid motion into (large-scale) oscillatory 
motion (oscillation modes), with displacement $\boldsymbol\xi(\vec{r},t)$, and
small-scale turbulent motion with the convective velocity field 
$\vec{u}(\vec{r},t)=(u,v,w)$. 
This separation is
superficially straightforward for radial p modes \citep{Gough77a}; however, for 
nonradial modes this separation is a much more difficult task
and can possibly not be accomplished for the very-high-degree p modes and
also very-high-order g modes \citep[see e.g.][]{Appourchaux09}. For simplicity
we shall discuss here only radial p modes.
\begin{figure}[t]
\centering
\includegraphics[width=0.48\textwidth]{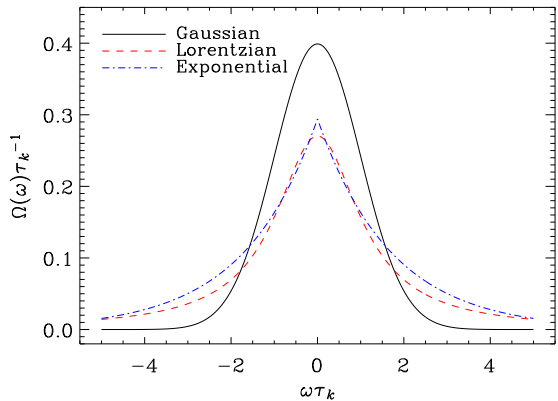}
\caption{Comparison of various analytical frequency factors 
$\Omega(\omega, \tau_k;\vec{r})$ 
[Eq.\,(\ref{e:Exponential})-(\ref{e:Lorentzian})] that are used in 
stochastic excitation models
}
\protect\label{fig:turb_spect1}
\end{figure}

Following the basic principle of Lighthill's acoustic analogy 
\citep[][see also \citealt{Crighton75}]{Lighthill52} one typically 
arranges the linearized global mode variables on the left and
the nonlinear fluctuating terms associated with the turbulent convection on
the right of the equations of motion, which may then be written as
\begin{equation}
\rho\left(\frac{\partial^2\boldsymbol\xi}{\partial t^2}
         +2\eta\frac{\partial\boldsymbol\xi}{\partial t}
         +\mathcal{L}\boldsymbol\xi\right)=
\boldsymbol{\mathcal{F}}(\vec{u}),
\label{e:excitation}
\end{equation}
in which the linear spatial differential operator $\cal L$ satisfies
the (unforced) homogeneous wave equation 
${\cal L}\boldsymbol\xi_i=\omega_i^2\boldsymbol\xi$
for the mode eigenfunctions $\boldsymbol\xi_i$ with
frequencies $\omega_i$, 
and $\boldsymbol{\mathcal{F}}$ is the nonlinear stochastic forcing and
damping term that depends only on the turbulent velocity field~$\vec{u}$.
Here we consider only the Reynolds stress driving term, which dominates
over the entropy driving term in the Sun and in most solar-type stars
\citep{Balmforth92b, SteinNordlund01, Stein04, Belkacem06, Samadi07}.
It should, however, be noted that there is partial cancellation between
the fluctuating Reynolds stress and entropy source terms, which could lead
to smaller oscillation amplitudes than adopting the Reynolds stress
contribution alone \citep{Osaki90, Stein04, Houdek06}.
Because the mode amplitudes of solar-type oscillations are small it is
additionally assumed that they do not interact with the nonlinear right
hand side of the wave equation, and consequently 
one can solve Eq.\,(\ref{e:excitation}) by a nonsingular perturbation 
method \citep{Goldreich77, Balmforth92b, Samadi01, Chaplin05}.
For radial modes only the vertical component $\mathcal{F}_r$ of the 
fluctuating Reynolds stress source term $\boldsymbol{\mathcal{F}}$ is 
important,
\begin{figure}[t]
\centering
\includegraphics[width=0.48\textwidth]{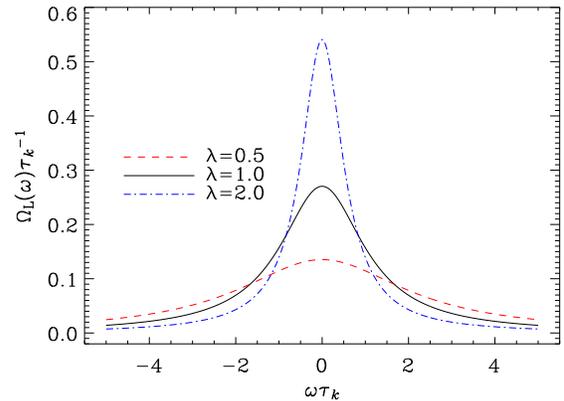}
\caption{Effect of varying the correlation parameter $\lambda$
[Eq.\,(\ref{e:correlation-time})] on the Lorentzian
frequency factor $\Omega_{\rm L}(\omega, \tau_k;\vec{r})$ 
[Eq.\,(\ref{e:Lorentzian})]
}
\protect\label{fig:turb_spect2}
\end{figure}
\begin{equation}
\mathcal{F}_r(w)=\partial_r(\rho ww-\langle\rho ww\rangle)\,,
\end{equation}
which depends on the $(r,r)$ component of a two-point correlation function, 
$R_{rr}:=\langle ww\rangle$, where angular brackets denote an ensemble average.
For incompressible, homogeneous isotropic turbulence the 
Fourier transform $\hat R_{rr}$ of $R_{rr}$ is proportional to the turbulent
energy spectrum function $E(k,\omega)$, i.e.
\begin{equation}
\hat R_{rr}\propto k^{-2}E(k,\omega)
\end{equation}
\citep{Batchelor53}, where $k$ is a wavenumber. 
\begin{figure}[t]
\centering
\includegraphics[width=0.48\textwidth]{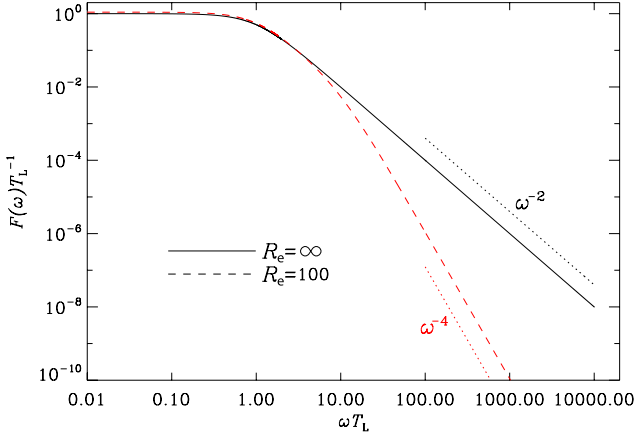}
\caption{Temporal spectrum of the normalized velocity autocorrelation 
(\ref{e:fourier2}) for two values of the 
Reynolds number $R_{\rm e}$, computed according to Sawford's (1991) 
second-order stochastic turbulence model 
}
\protect\label{fig:sawford}
\end{figure}
Following \citet{Stein67} we factorize the energy spectrum function into
$E(k,\omega)=E(k\ell/\pi)\Omega(\omega, \tau_k; r)$, 
where 
\begin{equation}
\tau_k=\lambda/ku_k
\label{e:correlation-time}
\end{equation}
is the correlation time-scale of eddies with vertical size $\ell=\pi/k$ and 
velocity $u_k$;
the correlation factor $\lambda$ is of order unity and accounts for
uncertainties in defining $\tau_k$.
For statistically stationary turbulence the energy supply rate is then 
given by (see e.g., Chaplin et~al. 2005 for details)
\begin{equation}
P=\frac{\pi}{9I}
  \int_0^R \ell^3
  \left(\Phi\Psi rp_{\rm t}\frac{\partial{\xi}_{ri}}{\partial r}\right)^2
  {\cal S}(\omega_i;r)\,{\rm d}r\,,
\label{e:energy-supply-rate}
\end{equation}
with
\begin{equation}
{\cal S}(\omega_i;r)=\int_0^\infty \kappa^{-2}\tilde E^2(\kappa)
                     \tilde\Omega(\omega_i,\tau_k;r)\,{\rm d}\kappa\,,
\label{e:function-S}
\end{equation}
where $p_{\rm t}=\langle\rho ww\rangle$ is the $(r,r)$ component of the 
(mean) Reynolds stress tensor, $\kappa=k\ell/\pi$, $R$ is surface radius,
$\xi_{ri}$ is the normalized radial part of $\boldsymbol{\xi}_i$, and the
product of $\Phi$ and $\Psi$ is a factor of unity accounting for the 
anisotropy of the turbulent velocity field.
The spectral function ${\cal S}$ accounts for contributions to $P$
from the small-scale turbulence. For the normalized spatial
turbulence energy spectrum $\tilde E(\kappa)$ it has been common to 
adopt, for example, the \citet{Kolmogorov41} spectrum. 
For the frequency-dependent factor $\Omega(\omega,\tau_k;r)$,
however, which is used for evaluating the self-convolution 
$\tilde\Omega(\omega_i,\tau_k;r)=
\int\Omega(\omega,\tau_k;r)\Omega(\omega_i\!-\!\omega,\tau_k;r)\,{\rm d}\omega$
in Eq.\,(\ref{e:function-S}),
no satisfactory theory exists. Various functional forms were proposed
in the past, which lead, however, to rather different values
for the modelled oscillation amplitudes $H$
\citep{Samadi03, Chaplin05, Appourchaux09}. We shall therefore discuss
the most commonly adopted frequency-dependent factors for stellar turbulence
spectra in the next section. 
\begin{figure}[t]
\centering
\includegraphics[width=0.36\textwidth]{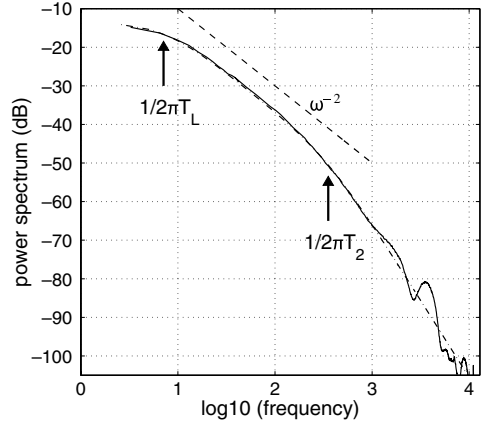}
\caption{Measured temporal velocity power spectrum in a rotating shear
turbulence experiment with water (solid curve). The dot-dashed curve is
a fit of Sawford's (1991) model, Eq.\,(\ref{e:fourier2}), to the data with
$R_{\rm e}=(T_{\rm L}/T_2)^2$ (adapted from \citealt{Mordant04})
}
\protect\label{fig:mordant}
\end{figure}
\vspace{-3pt}
\subsection{Turbulence spectra}
\label{ss:spectra}
\vspace{-3pt}
Differences in the (properly normalized) spatial spectrum $\tilde E(\kappa)$ 
create only minor differences in the energy supply rate $P$ 
\citep[e.g.,][]{Balmforth92b, Samadi01, Chaplin05}. But differences in the
frequency-dependent (temporal) spectrum $\Omega(\omega,\tau_k;r)$ would create 
rather large differences in the p-mode and also g-mode amplitudes 
\citep{Samadi03, Chaplin05, Belkacem09, Appourchaux09}. Those differences
are produced by the convection-oscillation interactions in the deeper
convectively unstable stellar layers, which are off resonance, and whose 
magnitude depends crucially on the adopted frequency dependence of the 
turbulence spectrum in the high-frequency tail of the cascade. 
We consider the following factors:\\

\noindent
(i) the Exponential factor \citep{Stein67}
\begin{equation}
\Omega_{\rm E}(\omega,\tau_k; r)=
    \frac{\sqrt{2\ln2}}{4}\tau_k\,{\rm e}^{-(\omega\tau_k\sqrt{2\ln2}/2)^2}\,,
\label{e:Exponential}
\end{equation}
(ii) the Gaussian factor \citep{Stein67}
\begin{equation}
\Omega_{\rm G}(\omega,\tau_k;r)=
    \frac{\tau_k}{\sqrt{2\pi}}\,{\rm e}^{-(\omega\tau_k/\sqrt2)^2}\,,
\label{e:Gaussian}
\end{equation}
(iii) the Lorentzian factor
   \citep{Gough77a, Samadi03, Chaplin05}
\begin{equation}
\Omega_{\rm L}(\omega,\tau_k;r)=
 \frac{\tau_k}{\pi\sqrt{2\ln2}}\,\frac{1}{1+(\omega\tau_k/\sqrt{2\ln2})^2}\,.
\label{e:Lorentzian}
\end{equation}
\begin{figure}[t]
\centering
\includegraphics[width=0.48\textwidth]{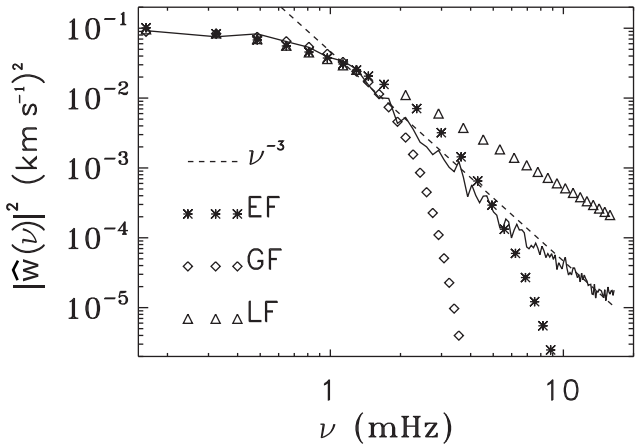}
\caption{Comparison of the temporal power spectrum of the vertical 
convective velocity $w$ in the Sun, computed with Stein \& Nordlund's 
3D hydrocode (solid curve), with analytical frequency factors (symbols). 
Results are shown for a depth of 250\,km below the surface and
for a horizontal wavenumber $k=4\,$Mm$^{-1}$ 
(from \citealt{Georgobiani04})
}
\protect\label{fig:dali1}
\end{figure}
Fig.\,\ref{fig:turb_spect1} compares the dimensionless quantity
$\Omega(\omega,\tau_k;r)\tau^{-1}_k$ for the three frequency factors
(\ref{e:Exponential})--(\ref{e:Lorentzian}). All three factors are normalized
according to $\int\Omega(\omega,\tau_k;r)\,{\rm d}\omega=1$, the integral
being from $-\infty$ to $\infty$. Furthermore, the full-width 
at half-maximum, i.e. $2\sqrt{2\ln2}\tau^{-1}_k$, has been chosen to be the 
same for all three factors, in order to have a meaningful comparison.
The Gaussian frequency factor decreases more rapidly with
$\omega\tau_k$, than both the Exponential and Lorentzian factors, i.e. 
the Gaussian factor produces the smallest magnitude in the high-frequency 
tail of the spectrum. Convective eddies situated in the deeper convectively
unstable stellar layers have longer characteristic time scales $\tau_k$, i.e.
$\tau_k$ [see Eq.\,(\ref{e:correlation-time})] increases with depth in the star
\citep[see for example Fig.~5 of][]{Chaplin05}. The Lorentzian factor 
decreases more slowly with depth at constant frequency and consequently a
larger fraction of the integrands of Eqs\,(\ref{e:energy-supply-rate}) and 
(\ref{e:function-S}) arises from large off-resonant eddies situated deep
in the star, whereas the Gaussian frequency factor gives less weight to the 
large off-resonance eddies. As a result of this the modelled oscillation
amplitudes are larger with a Lorentzian time-correlation function and 
therefore in better agreement with the observations than with
a Gaussian \citep{Samadi03, Samadi07}. However, \citet{Chaplin05} reported that
the Lorentzian time-correlation function leads to overestimated heights $H$
at low frequencies for solar p modes. Also for solar g modes
\citet{Belkacem09} concluded that the best fit for reproducing the observations
was obtained with a Gaussian factor at the lowest and a Lorentzian at the 
highest frequencies. Moreover, \citet{Belkacem09} reported that the correlation
parameter $\lambda$ [see Eq.\,(\ref{e:correlation-time})] has to increase with
stellar depth in order to reproduce the 3D numerical simulations by
\citet{Miesch08}, a result that is consistent with the findings by
\citet{Chaplin05}. The effect of $\lambda$ on the time-correlation functions
is demonstrated in Fig.\,\ref{fig:turb_spect2} for a Lorentzian factor.
Increasing $\lambda$ leads to a more rapidly declining tail and
consequently to a smaller contribution to the energy supply rate $P$ from 
the off-resonant eddies situated in the deep convection zone.
Finding the correct time-correlation function for stellar turbulence remains 
an open issue.
\begin{figure}[t]
\centering
\includegraphics[width=0.46\textwidth]{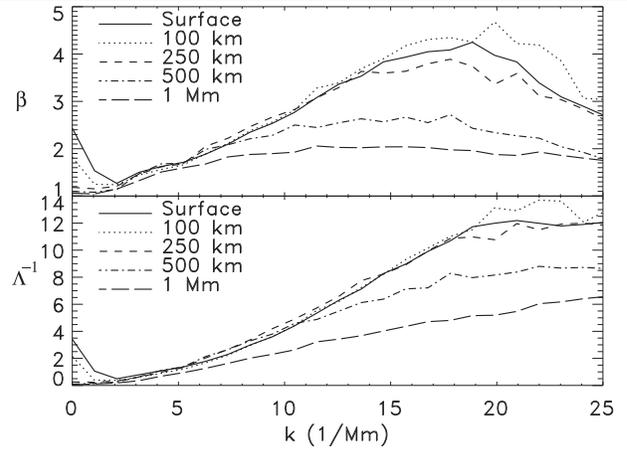}
\caption{Exponent $\beta$ and width $\Lambda^{-1}$ for the empirical 
expression (\ref{e:dali_fit}) for different (mean) horizontal wavenumbers 
$k$ and solar depths. $\beta$ \& $\Lambda$ are obtained from fitting
to the simulated temporal power spectra of the vertical velocity
the expression (\ref{e:dali_fit}) (from \citealt{Georgobiani04})
}
\protect\label{fig:dali2}
\end{figure}

From the theoretical point of view a Lorentzian factor is a result predicted
for the largest, most-energetic eddies by the time-dependent mixing-length
formulation of \citet{Gough77a}. Moreover, in agreement with
\citet{Kolmogorov41} theory \citet{Sawford91} demonstrated that in the 
limit of high-Reynolds-number turbulence the two-point velocity
autocorrelation, normalized by the velocity variance 
$\sigma^2:=\langle ww\rangle-\langle w\rangle\langle w\rangle$, is
\begin{equation}
\tilde R(\tau)=\langle w(t)w(t+\tau)\rangle/\sigma^2={\rm e}^{-\vert\tau\vert/T_{\rm L}}\,,
\label{e:autocorr}
\end{equation}
where 
\begin{equation}
T_{\rm L}=\int_0^t\tilde R(\tau)\,{\rm d}\tau
\end{equation}
is the Lagrangian integral time scale 
\citep[see also][]{Legg82}. Consequently the normalized velocity
autocorrelation $\tilde R$ is proportional to an exponential decrease with 
separation time $\tau$ in the inertial range, and the velocity spectrum
becomes with Eq.\,(\ref{e:autocorr})
\begin{eqnarray}
F(\omega)&=&\int_0^\infty\tilde R(\tau)\cos(\omega\tau)\,{\rm d}\tau
            \nonumber\\
         &=&\frac{T_{\rm L}}{1+(\omega T_{\rm L})^2}\,,
\label{e:fourier1}
\end{eqnarray}
i.e. it is of Lorentzian shape with $F(\omega)\propto\omega^{-2}$ for 
$\omega\rightarrow\infty$ and at infinite Reynolds number $R_{\rm e}$. Sawford
extended this analysis for arbitrary values of the Reynolds number
$R_{\rm e}:=(t_{\rm E}/t_\eta)^2$, where $t_{\rm E}$ and $t_\eta$ are
the time scales of the energy-containing eddies and the Kolmogorov
timescale respectively \citep{Tennekes72}. For this case the velocity 
spectrum is \citep{Sawford91}
\begin{equation}
F(\omega)=\frac{(1+R_{\rm e}^{1/2})R_{\rm e}^{1/2}T_{\rm L}}
{(R_{\rm e}^{1/2}-\omega^2T^2_{\rm L})^2+(1+R_{\rm e}^{1/2})^2\omega^2T^2_{\rm L}}\,.
\label{e:fourier2}
\end{equation}
Fig.\,\ref{fig:sawford} displays the spectrum $F(\omega)$ of the
normalized velocity autocorrelation $\tilde R$ for
two different values of the Reynolds number, $R_{\rm e}=(100,\infty)$. It 
is interesting to note that there is essentially no inertial range with
a frequency decay of $\omega^{-2}$, i.e. a Lorentzian frequency dependence
[Eq.\,(\ref{e:fourier1})], for $R_{\rm e}=100$; the spectrum decreases as 
$\omega^{-4}$ in the dissipation subrange, i.e. for turbulent scales close 
to the Kolmogorov dissipation scale.

These findings are supported by laboratory experiments. Fig.\,\ref{fig:mordant}
displays the measured velocity power spectrum in a rotating shear turbulence 
experiment (K{\`a}rm{\`a}n swirling flow) with water. The solid curve 
displays the measurements and the dot-dashed curve is a fit of 
expression\,(\ref{e:fourier2}) to the data with 
$R^{1/2}_{\rm e}=T_{\rm L}/T_2$, where 
$T_{\rm L}\simeq T_{\rm E}$ and $T_2\simeq t_\eta$. There is a well defined
inertial range for frequencies between $(2\pi T_{\rm L})^{-1}$ and
$(2\pi T_2)^{-1}$ with a decay of about $(\omega/2\pi)^{-2}$, i.e.
a Lorentzian spectrum. The high-frequency tail, however, decays more rapidly. 
The measurements are superficially similar to the second-order model
by \citet{Sawford91} depicted in Fig.\,\ref{fig:sawford}. It should, however, 
be noted that the molecular Prandtl number of about 6.8 in the water 
experiment is several magnitudes larger than the molecular Prandtl number
in stellar convection zones, which is typically of order $10^{-6}$.

\citet{Georgobiani04} used, as did \citet{Samadi03}, the 3D 
Large-Eddy-Simulations by \citet{SteinNordlund01} to investigate
the time-correlation function $\Omega(\omega, \tau_k; r)$ in the Sun.
Fig.\,\ref{fig:dali1} compares the power spectrum
(squared fast Fourier transform), $|w(\nu)|^2$, of the vertical 
component of the turbulent velocity field, ${\boldsymbol u}=(u,v,w)$, at 
fixed horizontal wavenumber $k$ and depth, between the simulation 
results (solid curve) and various analytical time-correlation functions 
[Exponential (EF), Gaussian (GF) and Lorentzian (LF) frequency factors]. 
The results are shown for a
wavenumber $k=4\,$Mm$^{-1}$ computed 250\,km below the solar surface.
Neither of the three analytical functions fit the simulation results,
particularly in the high-frequency tail of the spectrum.
Moreover, \citet{Georgobiani04} reported that the functional form (frequency 
dependence) of the simulation results changes substantially with wavenumber
and depth, and consequently that the turbulent energy spectrum 
function $E(k,\nu)$ is not separable into  wavenumber $E(k)$ and
frequency $\Omega(\nu, \tau_k; r)$. The authors suggest using instead the 
following empirical function for the temporal part
of the vertical velocity power spectrum
\begin{equation}
\vert\hat w(\nu)\vert^2\propto\frac{1}{\left(1+\Lambda^2\nu^2\right)^\beta}\,,
\label{e:dali_fit}
\end{equation}
where the two coefficients $\Lambda$ and $\beta$ are determined from
fitting Eq.\,(\ref{e:dali_fit}) to the 3D simulation results.
Note that for $\beta=1$ a Lorentzian frequency factor is recovered.
Results from solar simulations are given in 
Fig.\,\ref{fig:dali2} for various horizontal wavenumbers and depths. 
In agreement with 
\citet{Chaplin05} and \citet{Belkacem09} the coefficient 
$\Lambda\propto\lambda$ increases with depth, i.e. the width of the frequency 
factor decreases with depth (see also Fig.\,\ref{fig:turb_spect2}), 
particularly at smaller scales (i.e. at large wavenumbers). Except near 
the surface the values of the exponent $\beta$ are rather larger 
than unity.

\begin{figure}[t]
\centering
\includegraphics[width=0.45\textwidth]{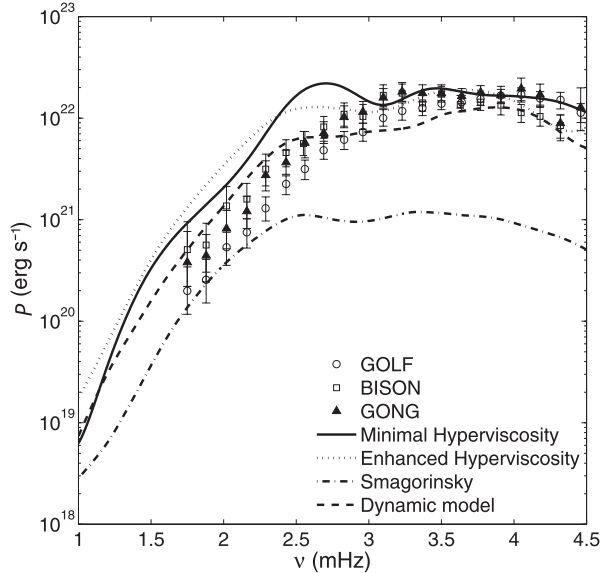}
\caption{Comparison of observed (symbols) and simulated (curves) stochastic
energy supply rates $P$ for solar p modes. Simulation results are shown
for different sub-grid models (from Jacoutot et al. 2008)
}
\protect\label{fig:jacoutot}
\end{figure}

\subsection{Effect of sub-grid models on simulation results}
\label{ss:sub-grid}
Numerical simulations of stellar convection resolve only the largest scales
of the turbulent motion. The smaller scales still need to be approximated
by a so-called sub-grid model. Such simulations are called 
Large-Eddy-Simulations. The total number of scales that are numerically
resolved, i.e. the ratio of the largest scale $l_{\rm L}$ to the smaller scale 
$l_\eta$, is related to the Reynolds number by \citep[e.g.][]{Tennekes72}
${l_{\rm L}}/{l_\eta}\sim R^{3/4}_{\rm e}\,,$
and consequently the total number $N$ of grid points in 
the simulation is
$N=({l_{\rm L}}/{l_\eta})^3\sim R^{9/4}_{\rm e}\,.$
A solar Reynolds number of $R_{\rm e}\simeq 10^{12}$
would require a total meshpoint number
of $N\simeq10^{27}$. With today's super computers the maximum achievable number 
of meshpoints is $N_{\rm max}\simeq10^{12}$, and is therefore some 
15 magnitudes too small for what is required to resolve all the turbulent 
scales of solar convection. Consequently a sub-grid model is necessary for
describing the dynamics of the numerically unresolved smaller scales of 
the turbulent cascade.
Various models are available. The most commonly used models are hyperviscosity 
models and the Smagorinsky model. All sub-grid models assume that turbulent
transport is a diffusive process. Hyperviscosity models, for example, use
higher derivatives for the diffusion operator in the momentum equation, 
thereby extending the inertial range, which also leads to a better 
representation of the dynamics of the larger scales. An overview of 
sub-grid models was recently presented by \citet{Miesch05}.\\
An obvious question to ask is what are the effects of using different
sub-grid models on the mode properties in simulations of solar-type stars? 
This question was recently addressed, in part, by \citet{Jacoutot08}, who 
compared simulated solar energy supply 
rates $P$ using various sub-grid models. Their results are summarized in 
Fig.\,\ref{fig:jacoutot}. Agreement with observations is generally
satisfactory except for the classical Smagorinsky model.

It is also important to note that the Prandtl number in 3D simulations
is currently about 0.01 $-$ 0.25 \citep[e.g.][]{Miesch08}. It is therefore
substantially larger than the Prandtl number in the Sun and in solar-type 
stars. Although 
numerical simulations provide an important input to our understanding of 
stellar convection and are very important for calibrating semi-analytical 
convection models, we must remain aware of the shortcomings of the currently
used 3D numerical simulations.
 
%
\acknowledgments
I am very grateful to Douglas Gough for many helpful discussions.
Support by the Austrian Science Fund (FWF project P21205) is 
thankfully acknowledged.

\makeatletter
\let\clear@thebibliography@page=\relax
\makeatother

\end{document}